\documentclass[preprint,prd,aps,showpacs,showkeys,nofootinbib]{revtex4}
\usepackage{graphicx}
\begin{document}

%\preprint{hep-ph/0412147}

\title{Lepton flavor violation in the $\mu\nu$SSM with slepton flavor mixing}

\author{Hai-Bin Zhang$^{a,b,}$\footnote{email:hbzhang@mail.dlut.edu.cn}, Tai-Fu Feng$^{a,b}$, Shu-Min Zhao$^b$, Fei Sun$^{a}$}

\affiliation{$^a$Department of Physics, Dalian University of Technology, Dalian, 116024, China\\
$^b$Department of Physics, Hebei University, Baoding, 071002, China}

\begin{abstract}
The $\mu\nu$SSM, one of supersymmetric extensions of the Standard Model, introduces three right-handed neutrino superfields to solve the $\mu$ problem and violates lepton number. Within framwork of the $\mu\nu$SSM, we investigate the lepton flavor violating (LFV) processes $Z\rightarrow l_i^\pm l_j^\mp$ with slepton flavor mixing. Simultaneously, we consider the LFV processes $l_j^-\rightarrow l_i^-\gamma$, $l_j^- \rightarrow l_i^- l_i^- l_i^+$ and muon conversion to electron in nuclei.
\end{abstract}


\maketitle

\section{Introduction\label{sec1}}
The observations of neutrino oscillations (see Refs.~\cite{neu-b1,neu-b2,neu-b3,neu-b4,neu-data}) imply that neutrinos have tiny masses and are mixed, which have demonstrated that lepton flavor in neutrino sector is not conserved. Nevertheless, in the Standard Model (SM) with massive neutrinos, the expected rates for the charged lepton flavor violating (LFV) processes are
very tiny, for instance ${\rm{Br}}(\mu\rightarrow e\gamma)<10^{-54}$~\cite{uer-SM1,uer-SM2} and ${\rm{Br}}(Z\rightarrow l_i^\pm l_j^\mp)\sim10^{-54}$~\cite{Zll-vSM1,Zll-vSM2,Zll-vSM3,Zll-vSM4}, which are far from the experimental reach. In Table~\ref{tab1}, we show the present experimental limits and future sensitivities for some LFV processes~\cite{Z-exp1,Z-exp2,Z-exp3,uer-exp1,uer-exp2,t-exp1,t-exp2,u3e-exp1,u3e-exp2,t-exp3,ueN-exp1,ueN-exp2}. Thus, any signal of LFV in charged lepton sector would be a hint of new physics.

\begin{table*}[htbp]
\caption{Present experimental limits and future sensitivities for some LFV processes.}
\begin{tabular*}{\textwidth}{@{\extracolsep{\fill}}llllll@{}}
\hline
LFV process & Present limit & Future sensitivity \\
\hline
$Z\rightarrow e\mu$ & $<1.7\times10^{-6}$ \cite{Z-exp1} & $\sim2.0\times10^{-9}$ \cite{Z-exp3}  \\
$Z\rightarrow e\tau$ & $<9.8\times10^{-6}$ \cite{Z-exp1} & $\sim(1.3-6.5)\times10^{-8}$ \cite{Z-exp3}  \\
$Z\rightarrow \mu\tau$ & $<1.2\times10^{-5}$ \cite{Z-exp2} & $\sim(0.44-2.2)\times10^{-8}$ \cite{Z-exp3}  \\
$\mu\rightarrow e\gamma$ & $<5.7\times10^{-13}$ \cite{uer-exp1} & $\sim6\times10^{-14}$ \cite{uer-exp2}  \\
$\tau\rightarrow e\gamma$ & $<3.3\times10^{-8}$ \cite{t-exp1} & $\sim10^{-8}-10^{-9}$ \cite{t-exp2}  \\
$\tau\rightarrow \mu\gamma$ & $<4.4\times10^{-8}$ \cite{t-exp1} & $\sim10^{-8}-10^{-9}$ \cite{t-exp2}  \\
$\mu\rightarrow 3e$ & $<1.0\times10^{-12}$ \cite{u3e-exp1} & $\sim10^{-16}$ \cite{u3e-exp2}  \\
$\tau\rightarrow 3e$ & $<2.7\times10^{-8}$ \cite{t-exp3} & $\sim10^{-9}-10^{-10}$ \cite{t-exp2}  \\
$\tau\rightarrow 3\mu$ & $<2.1\times10^{-8}$ \cite{t-exp3} & $\sim10^{-9}-10^{-10}$ \cite{t-exp2}  \\
$\mu\rightarrow e:\rm{Ti}$ & $<6.1\times10^{-13}$ \cite{ueN-exp1} & $\sim10^{-18}$ \cite{ueN-exp2}  \\
\hline
\end{tabular*}
\label{tab1}
\end{table*}

Several predictions for the LFV processes $Z\rightarrow l_i^\pm l_j^\mp$ have been obtained in the framework of various SM extensions~\cite{Zll-R1,Zll-R2,Zll-R3,Zll-R4,Zll-R5,Zll-R6,Zll-R7,Zll-R8,Zll-R9,Zll-R10,Zll-R11,Zll-R12,Zll-R13,Zll-R14,Zll-R15,Zll-R16,Zll-R17}. In this work, we investigate the processes $Z\rightarrow l_i^\pm l_j^\mp$ in the ``$\mu$ from $\nu$ Supersymmetric Standard Model'' ($\mu\nu$SSM)~\cite{mnSSM,mnSSM1,mnSSM2}. Within the $\mu\nu$SSM, nonzero vacuum expectative values (VEVs) of sneutrinos lead to R-parity and lepton number violations, and generate three tiny massive Majorana neutrinos at the tree level through the seesaw machanism~\cite{mnSSM,mnSSM1,mnSSM2,meu-m,meu-m1,meu-m2,meu-m3,neu-zhang1}. Especially, the $\mu$ problem~\cite{m-problem} of the Minimal Supersymmetric Standard Model (MSSM)~\cite{MSSM,MSSM1,MSSM2,MSSM3,MSSM4} had been solved in the $\mu\nu$SSM, via the R-parity breaking couplings ${\lambda _i}\hat \nu _i^c\hat H_d^a\hat H_u^b$ in the superpotential. The $\mu$ term is generated spontaneously through the nonzero VEVs of right-handed sneutrinos, $\mu  = {\lambda _i}\left\langle {\tilde \nu _i^c} \right\rangle$, when the electroweak symmetry is broken (EWSB).

In our previous work, we had analyzed some LFV processes $l_j^-\rightarrow l_i^-\gamma$, $l_j^-  \rightarrow l_i^- l_i^- l_i^+$ and  muon conversion to electron in nuclei within the $\mu\nu$SSM, under minimal flavor violation assumptions~\cite{ref-zhang,ref-zhang1}. The numerical results show that the expected rates for the LFV processes under minimal flavor violation assumptions, still remain orders of magnitude below the future experimental sensitivities. In this work, we continue to analyze the LFV processes with slepton flavor mixing, including $Z\rightarrow l_i^\pm l_j^\mp$. 

The paper has the following structure. In Section~\ref{sec2}, we present the $\mu\nu$SSM briefly, including its superpotential and the general soft SUSY-breaking terms. Section~\ref{sec3} contains the analytical expressions of the LFV processes $Z\rightarrow l_i^\pm l_j^\mp$. In Section~\ref{sec4}, we give the numerical analysis, under some assumptions and constraints on parameter space. The summary is given in Section~\ref{sec5}. The couplings are collected in Appendix~\ref{app-coupling}.

\section{the $\mu\nu$SSM\label{sec2}}

Besides the superfields of the MSSM, the $\mu\nu$SSM introduces three singlet right-handed neutrino superfields $\hat{\nu}_i^c\;(i=1,\;2,\;3)$. The corresponding superpotential of the $\mu\nu$SSM is given by~\cite{mnSSM}
\begin{eqnarray}
W =&&{\epsilon _{ab}} \Big( {Y_{u_{ij}}}\hat H_u^b\hat Q_i^a\hat u_j^c + {Y_{d_{ij}}}\hat H_d^a\hat Q_i^b\hat d_j^c
+ {Y_{e_{ij}}}\hat H_d^a\hat L_i^b\hat e_j^c + {Y_{\nu _{ij}}}\hat H_u^b\hat L_i^a\hat \nu _j^c \Big)  \nonumber\\
&&- {\epsilon _{ab}}{\lambda _i}\hat \nu _i^c\hat H_d^a\hat H_u^b + \frac{1}{3}{\kappa _{ijk}}\hat \nu _i^c\hat \nu _j^c\hat \nu _k^c \, ,
\end{eqnarray}
where $\hat H_d^T = \Big( {\hat H_d^0,\hat H_d^ - } \Big)$, $\hat H_u^T = \Big( {\hat H_u^ + ,\hat H_u^0} \Big)$, $\hat Q_i^T = \Big( {{{\hat u}_i},{{\hat d}_i}} \Big)$, $\hat L_i^T = \Big( {{{\hat \nu}_i},{{\hat e}_i}} \Big)$ are $SU(2)$ doublet superfields, and $\hat d_i^c$, $\hat u_i^c$ and $\hat e_i^c$ represent the singlet down-type quark, up-type quark
and charged lepton superfields, respectively. In addition, $Y_{u,d,\nu,e}$, $\lambda$, $\kappa$ respectively are dimensionless matrices, a vector, a totally symmetric tensor. And $a,b$ are SU(2) indices with antisymmetric tensor $\epsilon_{12}=-\epsilon_{21}=1$. In this paper, the summation convention is implied on repeated indices.

In the superpotential, the first three terms are the same as the MSSM. Next two terms can generate the effective bilinear terms $\epsilon _{ab} \varepsilon_i \hat H_u^b\hat L_i^a$, $\epsilon _{ab} \mu \hat H_d^a\hat H_u^b$,  and $\varepsilon_i= Y_{\nu _{ij}} \left\langle {\tilde \nu _j^c} \right\rangle$, $\mu  = {\lambda _i}\left\langle {\tilde \nu _i^c} \right\rangle$,  once the electroweak symmetry is broken. The last term generates the effective Majorana masses for neutrinos at the electroweak scale. And the last two terms explicitly violate lepton number and R-parity.

In the framework of supergravity mediated supersymmetry breaking, the general soft SUSY-breaking terms in the $\mu\nu$SSM are given as
\begin{eqnarray}
- \mathcal{L}_{soft}=&&m_{{{\tilde Q}_{ij}}}^{\rm{2}}\tilde Q{_i^{a\ast}}\tilde Q_j^a
+ m_{\tilde u_{ij}^c}^{\rm{2}}\tilde u{_i^{c\ast}}\tilde u_j^c + m_{\tilde d_{ij}^c}^2\tilde d{_i^{c\ast}}\tilde d_j^c
+ m_{{{\tilde L}_{ij}}}^2\tilde L_i^{a\ast}\tilde L_j^a  \nonumber\\
&&+ \; m_{\tilde e_{ij}^c}^2\tilde e{_i^{c\ast}}\tilde e_j^c + m_{{H_d}}^{\rm{2}} H_d^{a\ast} H_d^a
+ m_{{H_u}}^2H{_u^{a\ast}}H_u^a + m_{\tilde \nu_{ij}^c}^2\tilde \nu{_i^{c\ast}}\tilde \nu_j^c \nonumber\\
&&+ \; \epsilon_{ab} \Big[{{({A_u}{Y_u})}_{ij}}H_u^b\tilde Q_i^a\tilde u_j^c
+ {{({A_d}{Y_d})}_{ij}}H_d^a\tilde Q_i^b\tilde d_j^c + {{({A_e}{Y_e})}_{ij}}H_d^a\tilde L_i^b\tilde e_j^c + \textrm{H.c.} \Big] \nonumber\\
&&+ \; \Big[ {\epsilon _{ab}}{{({A_\nu}{Y_\nu})}_{ij}}H_u^b\tilde L_i^a\tilde \nu_j^c
- {\epsilon _{ab}}{{({A_\lambda }\lambda )}_i}\tilde \nu_i^c H_d^a H_u^b
+ \frac{1}{3}{{({A_\kappa }\kappa )}_{ijk}}\tilde \nu_i^c\tilde \nu_j^c\tilde \nu_k^c + \textrm{H.c.} \Big] \nonumber\\
&&- \; \frac{1}{2}\Big({M_3}{{\tilde \lambda }_3}{{\tilde \lambda }_3}
+ {M_2}{{\tilde \lambda }_2}{{\tilde \lambda }_2} + {M_1}{{\tilde \lambda }_1}{{\tilde \lambda }_1} + \textrm{H.c.} \Big)\,.
\end{eqnarray}
Here, the first two lines consist of mass squared terms of squarks, sleptons and Higgses. The next two lines contain the trilinear scalar couplings. In the last lines, $M_3$, $M_2$ and $M_1$ denote Majorana masses corresponding to $SU(3)$, $SU(2)$ and $U(1)$ gauginos $\hat{\lambda}_3$, $\hat{\lambda}_2$ and $\hat{\lambda}_1$, respectively.
In addition to the terms from $\mathcal{L}_{soft}$, the tree-level scalar potential receives the usual D and F term contributions~\cite{mnSSM1}.

Once the electroweak symmetry is spontaneously broken, the neutral scalars develop in general the following VEVs:
\begin{eqnarray}
\langle H_d^0 \rangle = \upsilon_d \,, \qquad \langle H_u^0 \rangle = \upsilon_u \,, \qquad
\langle \tilde \nu_i \rangle = \upsilon_{\nu_i} \,, \qquad \langle \tilde \nu_i^c \rangle = \upsilon_{\nu_i^c} \,.
\end{eqnarray}
Thus one can define neutral scalars as usual
\begin{eqnarray}
&&H_d^0=\frac{h_d + i P_d}{\sqrt{2}} + \upsilon_d, \qquad\; \tilde \nu_i = \frac{(\tilde \nu_i)^\Re + i (\tilde \nu_i)^\Im}{\sqrt{2}} + \upsilon_{\nu_i},  \nonumber\\
&&H_u^0=\frac{h_u + i P_u}{\sqrt{2}} + \upsilon_u, \qquad \tilde \nu_i^c = \frac{(\tilde \nu_i^c)^\Re + i (\tilde \nu_i^c)^\Im}{\sqrt{2}} + \upsilon_{\nu_i^c}.
\end{eqnarray}
And one can define
\begin{eqnarray}
\tan\beta=\frac{\upsilon_u}{\sqrt{\upsilon_d^2+\upsilon_{\nu_i}\upsilon_{\nu_i}}}.
\end{eqnarray}

For simplicity, we will assume that all parameters in the potential are real in the model. The CP-odd neutral scalar mass matrix $M_P^2$ and charged scalar mass matrix $M_{S^{\pm}}^2$ can respectively isolate massless unphysical Goldstone bosons $G^0$ and $G^{\pm}$, which can be written as~\cite{ref-zhang,ref-zhang1,ref-zhang2}
\begin{eqnarray}
&&G^0 = {1 \over \sqrt{\upsilon_d^2+\upsilon_u^2+\upsilon_{\nu_i} \upsilon_{\nu_i}}} \Big(\upsilon_d {P_d}-\upsilon_u{P_u}+\upsilon_{\nu_i}{(\tilde \nu_i)^\Im}\Big),\\
&&G^{\pm} = {1 \over \sqrt{\upsilon_d^2+\upsilon_u^2+\upsilon_{\nu_i} \upsilon_{\nu_i}}} \Big(\upsilon_d H_d^{\pm} - \upsilon_u {H_u^{\pm}}+\upsilon_{\nu_i}\tilde e_{L_i}^{\pm}\Big).
\end{eqnarray}
In the physical gauge, the Goldstone bosons $G^0$ and $G^{\pm}$ are eaten by $Z$-boson and $W$-boson, respectively, and disappear from the Lagrangian.
The masses of neutral and charged gauge bosons can be given by
\begin{eqnarray}
&&m_{Z}={e\over {\sqrt{2}s_{_W} c_{_W}}}\sqrt{\upsilon_u^2+\upsilon_d^2+\upsilon_{\nu_i} \upsilon_{\nu_i}},  \\ 
&&m_{W}={e\over\sqrt{2}s_{_W}}\sqrt{\upsilon_u^2+\upsilon_d^2+\upsilon_{\nu_i} \upsilon_{\nu_i}},
\end{eqnarray}
where $e$ denotes the electromagnetic coupling constant, $s_{_W}=\sin\theta_{_W}$ and $c_{_W}=\cos\theta_{_W}$ with the Weinberg angle $\theta_{_W}$, respectively.

\section{The LFV decays $Z\rightarrow l_i^\pm l_j^\mp$\label{sec3}}

\begin{figure}[htbp]
\setlength{\unitlength}{1mm}
\centering
\includegraphics[width=2.8in]{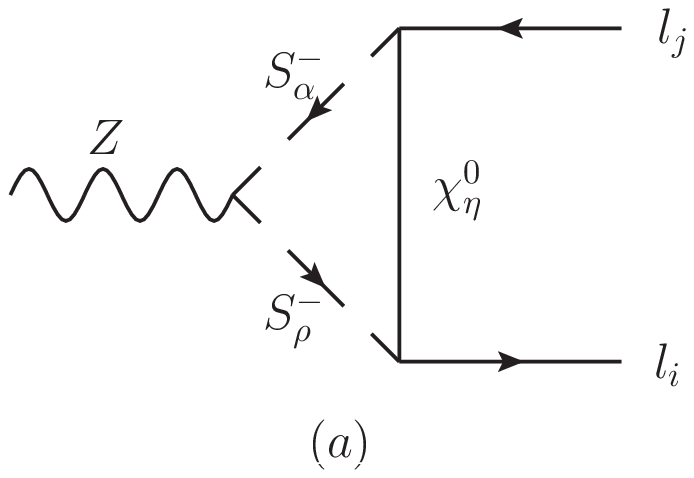}%
\includegraphics[width=2.8in]{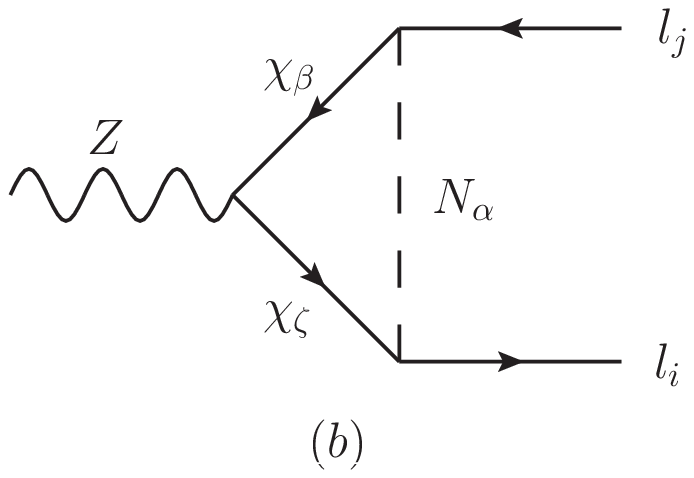}
\caption[]{Feynman diagrams for $Z\rightarrow l_i^\pm l_j^\mp$ in the $\mu\nu$SSM. (a) represents the contributions from neutral fermions $\chi_\eta^0$ and charged scalars $S_{\alpha,\rho}^-$ loops, while (b) represents the contributions from charged fermions $\chi_{\beta,\zeta}$ and neutral scalars $N_\alpha$ ($N=S,P$) loops.}
\label{fig1}
\end{figure}

The Feynman diagrams for $Z\rightarrow l_i^\pm l_j^\mp$ in the $\mu\nu$SSM are depicted by Fig.~\ref{fig1}. And the corresponding effective amplitude for $Z\rightarrow l_i^\pm l_j^\mp$ can be written as~\cite{Zll-R9}
\begin{eqnarray}
\mathcal{M}_{\mu}= e{\bar l_i}{\gamma _\mu }({F_L^{ij}}{P_L} + {F_R^{ij}}{P_R}){l_j},
\end{eqnarray}
with
\begin{eqnarray}
{F_{L,R}^{ij}} = F_{L,R}^{(n)ij} + F_{L,R}^{(c)ij},
\end{eqnarray}
where $F_{L,R}^{(n)ij}$ denote the contributions from the virtual neutral fermion loops, and $F_{L,R}^{(c)ij}$ stand for the contributions from the virtual charged fermion loops, respectively. After integrating the heavy freedoms out, we formulate those coefficients as follows:
\begin{eqnarray}
&&F_L^{(n)ij} = \sum\limits_{N=S,P} \Big[ \frac{{m_{{\chi _\zeta }}}{m_{{\chi _\beta }}}}
{{e}{m_W^2}}C_R^{{N_\alpha }{\chi _\zeta }{{\bar \chi }_{2+i}}}C_L^{Z{\chi _\beta }{{\bar \chi }_\zeta }}
C_L^{{N_\alpha }{\chi _{2+j}}{{\bar \chi }_\beta }}{G_1}({x_{{N_\alpha }}},{x_{{\chi _\zeta }}},{x_{{\chi _\beta }}})\nonumber\\
&&\qquad\qquad\qquad\quad - \: \frac{1}{2{e}} C_R^{{N_\alpha }{\chi _\zeta }{{\bar \chi }_{2+i}}}C_R^{Z{\chi _\beta }
{{\bar \chi }_\zeta }}C_L^{{N_\alpha }{\chi _{2+j}}{{\bar \chi }_\beta }}{G_2}({x_{{N_\alpha }}},
{x_{{\chi _\zeta }}},{x_{{\chi _\beta }}}) \Big],\\
&&F_L^{(c)ij} = \, \frac{1}{2{e}}C_R^{S_\rho ^ - \chi _\eta ^0{{\bar \chi }_{2+i}}}
C_R^{ZS_\alpha ^ - S_\rho ^ {-\ast} }C_L^{S_\alpha ^{-\ast} {\chi _{2+j}}\bar \chi _\eta ^0}{G_2}
({x_{\chi _\eta ^0}},{x_{S_\alpha ^ - }},{x_{S_\rho ^ - }}) ,\\
&&F_R^{(n,c)ij} = \left. {F_L^{(n,c)ij}} \right|{ _{L \leftrightarrow R}} .
\end{eqnarray}
Here, the concrete expressions for coupling coefficients $C_{L,R}$ can be found in Appendix~\ref{app-coupling}, $x= {m^2}/{m_W^2}$ and $m$ is the mass for the corresponding particle. And the form factors $G_{i} $ are given by
\begin{eqnarray}
&&{G_1}(\textit{x}_1 , x_2 , x_3) =  \frac{1}{{16{\pi ^2}}}\Big[ \frac{{{x_1}\ln {x_1}}}{{({x_1} - {x_2})({x_1} - {x_3})}} + \frac{{{x_2}\ln {x_2}}}{{({x_2} - {x_1})({x_2} - {x_3})}} \nonumber\\
&&\qquad\qquad\qquad\qquad\quad\;\;  + \: \frac{{{x_3}\ln {x_3}}}{{({x_3} - {x_1})({x_3} - {x_2})}}\Big], \\
&&{G_2}(\textit{x}_1 , x_2 , x_3) =  \frac{1}{{16{\pi ^2}}}\Big[  \frac{{x_1^2\ln {x_1}}}{{({x_1} - {x_2})({x_1} - {x_3})}}  + \frac{{x_2^2\ln {x_2}}}{{({x_2} - {x_1})({x_2} - {x_3})}}\nonumber\\
&&\qquad\qquad\qquad\qquad\quad\;\;  + \: \frac{{x_3^2\ln {x_3}}}{{({x_3} - {x_1})({x_3} - {x_2})}} \Big].
\end{eqnarray}

Then, we can obtain the branching ratio of $Z\rightarrow l_i^\pm l_j^\mp$ 
\begin{eqnarray}
{\rm{Br}}(Z\rightarrow l_i^\pm l_j^\mp)= \frac{e^2}{12\pi}\frac{m_Z}{{{\Gamma}}_Z}\Big({\left| {F_L^{ij}} \right|^2} + {\left| {F_R^{ij}} \right|^2}\Big),
\end{eqnarray}
where ${{\Gamma}}_Z$ denotes the total decay width of $Z$-boson. In the numerical calculation, we choose ${{\Gamma}}_Z \simeq 2.4952\:{\rm{GeV}}$~\cite{PDG}.

\section{Numerical analysis\label{sec4}}

In order to obtain a more transparent numerical results, we take the minimal flavor violation (MFV) assumptions for some parameters in the $\mu\nu{\rm SSM}$, which assume
\begin{eqnarray}
&&\;\,{\kappa _{ijk}} = \kappa {\delta _{ij}}{\delta _{jk}}, \;\;
{({A_\kappa }\kappa )_{ijk}} =
{A_\kappa }\kappa {\delta _{ij}}{\delta _{jk}}, \quad\;
\lambda _i = \lambda , \qquad\;\;\;\;
{{\rm{(}}{A_\lambda }\lambda {\rm{)}}_i} = {A_\lambda }\lambda,\nonumber\\
&&\;\,{Y_{{u _{ij}}}} = {Y_{{u _i}}}{\delta _{ij}},\quad
 (A_u Y_u)_{ij}={A_{u_i}}{Y_{{u_i}}}{\delta _{ij}},\;\;\:
{Y_{{\nu _{ij}}}} = {Y_{{\nu _i}}}{\delta _{ij}},\;\;\,
(A_\nu Y_\nu)_{ij}={a_{{\nu_i}}}{\delta _{ij}},\nonumber\\
&&\;\,{Y_{{d_{ij}}}} = {Y_{{d_i}}}{\delta _{ij}},\quad\:
(A_d Y_d)_{ij}={A_{d_i}}{Y_{{d_i}}}{\delta _{ij}},\quad
{Y_{{e_{ij}}}} = {Y_{{e_i}}}{\delta _{ij}},\qquad\;\;\;\,
\upsilon_{\nu_i^c}=\upsilon_{\nu^c},\nonumber\\
&&m_{\tilde Q_{ij}}^2 = m_{{{\tilde Q_i}}}^2{\delta _{ij}}, \qquad\:\,
m_{\tilde u_{ij}^c}^2 = m_{{{\tilde u_i}^c}}^2{\delta _{ij}}, \quad\;\:\:
m_{\tilde d_{ij}^c}^2 = m_{{{\tilde d_i}^c}}^2{\delta _{ij}}, \quad\;\;\;\;
m_{\tilde \nu_{ij}^c}^2 = m_{{{\tilde \nu_i}^c}}^2{\delta _{ij}},
\label{MFV}
\end{eqnarray}
where $i,\;j,\;k =1,\;2,\;3 $. Restrained by the quark and lepton masses, we could have
\begin{eqnarray}
{Y_{{u_i}}} = \frac{{{m_{{u_i}}}}}{{{\upsilon_u}}},\qquad 
{Y_{{d_i}}} = \frac{{{m_{{d_i}}}}}{{{\upsilon_d}}},\qquad 
{Y_{{e_i}}} = \frac{{{m_{{l_i}}}}}{{{\upsilon_d}}},
\end{eqnarray}
where $m_{u_i}$, $m_{d_i}$ and $m_{l_i}$ are the up-quark, down-quark and charged lepton masses, respectively, and we choose the values from Ref.~\cite{PDG}.

For soft breaking slepton mass matrices $m_{{{\tilde L},{\tilde e^c}}}^2$ and trilinear coupling matrix $({A_e}{Y_e})$, we will introduce the slepton flavor mixings, which take into account the off-diagonal terms for the matrices and are defined as~\cite{sl-mix,sl-mix1,sl-mix2,sl-mix3,sl-mix4,neu-zhang2}
\begin{eqnarray}
&&\quad\;\,{m_{\tilde L}^2} = \left( {\begin{array}{*{20}{c}}
   1 & \delta_{12}^{LL} & \delta_{13}^{LL}  \\
   \delta_{12}^{LL} & 1 & \delta_{23}^{LL}  \\
   \delta_{13}^{LL} & \delta_{23}^{LL} & 1  \\
\end{array}} \right){m_{L}^2},\\
&&\quad\:{m_{\tilde e^c}^2} = \left( {\begin{array}{*{20}{c}}
   1 & \delta_{12}^{RR} & \delta_{13}^{RR}  \\
   \delta_{12}^{RR} & 1 & \delta_{23}^{RR}  \\
   \delta_{13}^{RR} & \delta_{23}^{RR} & 1  \\
\end{array}} \right){m_{E}^2},\\
&&({A_e}{Y_e}) = \left( {\begin{array}{*{20}{c}}
   m_{l_1}{A_e} & \delta_{12}^{LR}{m_{L}}{m_{E}} & \delta_{13}^{LR}{m_{L}}{m_{E}}  \\
   \delta_{12}^{LR}{m_{L}}{m_{E}} & m_{l_2}{A_e} & \delta_{23}^{LR}{m_{L}}{m_{E}}  \\
   \delta_{13}^{LR}{m_{L}}{m_{E}} & \delta_{23}^{LR}{m_{L}}{m_{E}} & m_{l_3}{A_e}  \\
\end{array}} \right){1\over {\upsilon_d}}.
\end{eqnarray}
For simplicity, we will choose the slepton mixing parameters
\begin{eqnarray}
&&\delta_{12}^{LL}=\delta_{12}^{RR}=\delta_{12}^{LR}\equiv \delta_{12}^X, \nonumber\\ &&\delta_{13}^{LL}=\delta_{13}^{RR}=\delta_{13}^{LR}\equiv \delta_{13}^X, \nonumber\\ &&\delta_{23}^{LL}=\delta_{23}^{RR}=\delta_{23}^{LR}\equiv \delta_{23}^X.
\end{eqnarray}

At the EW scale, the soft masses $m_{\tilde H_d}^2$, $m_{\tilde H_u}^2$ and $m_{\tilde \nu_i^c}^2$ can be derived from the minimization conditions of the tree-level neutral scalar potential, which are given in Refs.~\cite{mnSSM1,ref-zhang}. Ignoring the terms of the second order in $Y_{\nu}$ and assuming $(\upsilon_{\nu_i}^2+\upsilon_d^2-\upsilon_u^2)\approx (\upsilon_d^2-\upsilon_u^2)$, one can have the minimization conditions of the tree-level neutral scalar potential with respect to $\upsilon_{\nu_i}\:(i=1,2,3)$ below~\cite{neu-zhang2}
\begin{eqnarray}
m_{\tilde L_{ij}}^2 \upsilon_{\nu_j}+{G^2\over 4} (\upsilon_d^2-\upsilon_u^2)\upsilon_{\nu_i}=\Big[\lambda \upsilon_d (\upsilon_u^2+\upsilon_{\nu^c}^2) - \kappa \upsilon_u \upsilon_{\nu^c}^2\Big] Y_{\nu_i} -\upsilon_u \upsilon_{\nu^c}a_{\nu_i},
\label{eq-min}
\end{eqnarray}
where $G^2=g_1^2+g_2^2$ and $g_1 c_{_W} =g_2 s_{_W}=e$. Solving Eq.~(\ref{eq-min}), we can gain the left-handed sneutrino VEVs
\begin{eqnarray}
\upsilon_{\nu_i}=\frac{{\rm{det}}\: T_i}{{\rm{det}}\: T},\qquad  (i=1,2,3),
\label{eq-vi}
\end{eqnarray}
where
\begin{eqnarray}
T = \left( {\begin{array}{*{20}{c}}
   m_{\tilde L_{11}}^2 +{G^2\over 4}(\upsilon_d^2-\upsilon_u^2) & m_{\tilde L_{12}}^2 & m_{\tilde L_{13}}^2  \\ [6pt]
   m_{\tilde L_{21}}^2 & m_{\tilde L_{22}}^2 +{G^2\over 4}(\upsilon_d^2-\upsilon_u^2) & m_{\tilde L_{23}}^2  \\ [6pt]
   m_{\tilde L_{31}}^2 & m_{\tilde L_{32}}^2 & m_{\tilde L_{33}}^2 +{G^2\over 4}(\upsilon_d^2-\upsilon_u^2)  \\ [6pt]
\end{array}} \right),
\end{eqnarray}
and $T_i$ can be acquired from $T$ by replacing the $i$-th column with
\begin{eqnarray}
\left( {\begin{array}{*{20}{c}}
   \Big[\lambda \upsilon_d (\upsilon_u^2+\upsilon_{\nu^c}^2) - \kappa \upsilon_u \upsilon_{\nu^c}^2\Big] Y_{\nu_1} -\upsilon_u \upsilon_{\nu^c}a_{\nu_1}  \\ [6pt]
   \Big[\lambda \upsilon_d (\upsilon_u^2+\upsilon_{\nu^c}^2) - \kappa \upsilon_u \upsilon_{\nu^c}^2\Big] Y_{\nu_2} -\upsilon_u \upsilon_{\nu^c}a_{\nu_2}  \\ [6pt]
   \Big[\lambda \upsilon_d (\upsilon_u^2+\upsilon_{\nu^c}^2) - \kappa \upsilon_u \upsilon_{\nu^c}^2\Big] Y_{\nu_3} -\upsilon_u \upsilon_{\nu^c}a_{\nu_3}  \\ [6pt]
\end{array}} \right).
\end{eqnarray}

In the $\mu\nu{\rm SSM}$, the sneutrino sector may appear the tachyons, which masses squared are negative. So, we need analyse the masses of the sneutrinos. The masses squared of left-handed sneutrinos are basically determined by soft breaking slepton mass matrix $m_{\tilde L}^2$. And the CP-even and CP-odd right-handed sneutrino masses squared can be approximately written as
\begin{eqnarray}
&&m_{S_{5+i}}^2\approx (A_\kappa+4\kappa\upsilon_{\nu^c})\kappa\upsilon_{\nu^c} +A_\lambda \lambda \upsilon_d \upsilon_u/\upsilon_{\nu^c}-2\lambda^2(\upsilon_d^2+\upsilon_u^2),\\
&&m_{P_{5+i}}^2\approx -3A_\kappa \kappa\upsilon_{\nu^c} +(A_\lambda/\upsilon_{\nu^c}+4\kappa)\lambda \upsilon_d \upsilon_u-2\lambda^2(\upsilon_d^2+\upsilon_u^2).
\end{eqnarray}
Here, the main contribution for the mass squared is the first term as $\kappa$ is large, in the limit of $\upsilon_{\nu^c} \gg \upsilon_{u,d}$. Therefore, we could use the approximate relation
\begin{eqnarray}
-4\kappa\upsilon_{\nu^c}\lesssim A_\kappa \lesssim 0,
\label{tachyon}
\end{eqnarray}
to avoid the tachyons.

Before calculation, the constraints on the parameters of the $\mu\nu{\rm SSM}$ from neutrino experiments should be considered at first. Three flavor neutrinos $\nu_{e,\mu,\tau}$ could mix into three massive neutrinos $\nu_{1,2,3}$ during their flight, and the mixings are described by the Pontecorvo-Maki-Nakagawa-Sakata unitary matrix $U_{_{PMNS}}$ \cite{ref-PMNS1,ref-PMNS2}. The experimental observations of the parameters in $U_{_{PMNS}}$ for the normal mass hierarchy show that \cite{neu-data}
\begin{eqnarray}
&&\sin^2\theta_{12} =0.302_{-0.012}^{+0.013},\qquad  \Delta m_{21}^2 =7.50_{-0.19}^{+0.18}\times 10^{-5} {\rm eV}^2,  \nonumber\\
&&\sin^2\theta_{23}=0.413_{-0.025}^{+0.037},\qquad  \Delta m_{31}^2 =2.473_{-0.067}^{+0.070}\times 10^{-3} {\rm eV}^2,  \nonumber\\
&&\sin^2 \theta_{13} =0.0227_{-0.0024}^{+0.0023}.
\label{neutrino-oscillation}
\end{eqnarray}

In the $\mu\nu{\rm SSM}$, the three tiny neutrino masses are obtained through TeV scale seesaw mechanism~\cite{mnSSM,mnSSM1,mnSSM2,meu-m,meu-m1,meu-m2,meu-m3,neu-zhang1}. Assumed that the charged lepton mass matrix in the flavor basis is in the diagonal form, we parameterize the unitary matrix which diagonalizes the effective light neutrino mass matrix $m_{eff}$ (see Ref.~\cite{ref-zhang}) as \cite{ref-Uv1,ref-Uv2}
\begin{eqnarray}
{U_\nu} = &&\left( {\begin{array}{*{20}{c}}
   {{c_{12}}{c_{13}}} & {{s_{12}}{c_{13}}} & {{s_{13}}{e^{ - i\delta }}}  \\
   { - {s_{12}}{c_{23}} - {c_{12}}{s_{23}}{s_{13}}{e^{i\delta }}} & {{c_{12}}{c_{23}} - {s_{12}}
   {s_{23}}{s_{13}}{e^{i\delta }}} & {{s_{23}}{c_{13}}}  \\
   {{s_{12}}{s_{23}} - {c_{12}}{c_{23}}{s_{13}}{e^{i\delta }}} & { - {c_{12}}{s_{23}} - {s_{12}}
   {c_{23}}{s_{13}}{e^{i\delta }}} & {{c_{23}}{c_{13}}}  \\
\end{array}} \right)  \nonumber\\
&&\times \: diag(1,{e^{i\frac{{{\alpha _{21}}}}{2}}},{e^{i\frac{{{\alpha _{31}}}}{2}}})\:,
\label{PMNS-matrix}
\end{eqnarray}
where ${c_{_{ij}}} = \cos {\theta _{ij}}$, ${s_{_{ij}}} = \sin {\theta _{ij}}$. In our calculation, the values of $\theta_{ij}$ are obtained from the experimental data in Eq.~(\ref{neutrino-oscillation}), and all CP violating phases $\delta$, $\alpha _{21}$ and $\alpha _{31}$ are set to zero. $U_\nu$ diagonalizes $m_{eff}$ in the following way:
\begin{eqnarray}
U_\nu ^T m_{eff}^T{m_{eff}}{U_\nu} = diag({m_{\nu _1}^2},{m_{\nu _2}^2},{m_{\nu _3}^2}).
\label{eff}
\end{eqnarray}
For the neutrino mass spectrum, we assume it to be normal hierarchical, i.e., ${m_{\nu_1}}{\rm{ < }}{m_{\nu_2}}{\rm{ < }}{m_{\nu_3}}$, and we choose the neutrino mass $m_{\nu_1}=10^{-2}\:{\rm{eV}}$ as input in our numerical analysis, limited on neutrino masses from neutrinoless double-$\beta$ decay~\cite{neu-m-limit} and cosmology~\cite{neu-m-limit1}. The other two neutrino masses $m_{\nu_{2,3}}$ can be obtained through the experimental data on the differences of neutrino mass squared in Eq.~(\ref{neutrino-oscillation}).  Then, we can numerically derive $Y_{\nu_i} \sim \mathcal{O}(10^{-7})$ and $a_{\nu_i} \sim \mathcal{O}(-10^{-4}{\rm{GeV}})$ from Eq.~(\ref{eff}). Accordingly, $\upsilon_{\nu_i} \sim \mathcal{O}(10^{-4}{\rm{GeV}})$ through Eq.~(\ref{eq-min}). Due to $\upsilon_{\nu_i}\ll\upsilon_{u,d}$, we can have
\begin{eqnarray}
\tan\beta\simeq \frac{\upsilon_u}{\upsilon_d}.
\end{eqnarray}

Recently, a neutral Higgs with mass around $125\;{\rm GeV}$ reported by ATLAS~\cite{ATLAS} and CMS~\cite{CMS} also contributes a strict constraint on relevant parameter space of the model. The global fit to the ATLAS and CMS Higgs data gives~\cite{mh-AC}:
\begin{eqnarray}
m_{{h}}=125.66\pm0.34\;{\rm GeV}.
\label{M-h}
\end{eqnarray}
In the $\mu\nu$SSM, the loop effects of right-handed neutrino/sneutrino on the SM-like Higgs mass can be neglected, due to small neutrino Yukawa couplings $Y_{\nu_i} \sim \mathcal{O}(10^{-7})$ and left-handed neutrino superfield VEVs $\upsilon_{\nu_i} \sim \mathcal{O}(10^{-4}{\rm{GeV}})$. Through the numerical computation in Ref.~\cite{ref-zhang}, we also can numerically ignore the radiative corrections from $b$ quark, $\tau$ lepton and their supersymmetric partners on the SM-like Higgs mass. Then, the main radiative corrections on the SM-like Higgs mass in the $\mu\nu$SSM come from the top quark and its supersymmetric partners, similarly to the MSSM. However when $\tan\beta$ is large enough, we also need to consider the radiative corrections from $b$ quark and its supersymmetric partners. Due to the introduction of some new couplings in the superpotential, the SM-like Higgs mass in the $\mu\nu$SSM gets additional contribution at tree-level~\cite{mnSSM1}. Therefore, the SM-like Higgs in the $\mu\nu{\rm SSM}$ can easily account for the mass around $125\,{\rm GeV}$, especially for small $\tan\beta$. For moderate $\tan\beta$ and large mass of the pseudoscalar $M_A$, the SM-like Higgs mass in the $\mu\nu{\rm SSM}$ is approximately given by
\begin{eqnarray}
m_h^2 \simeq m_Z^2 \cos^2 2\beta + \frac{6 \lambda^2 s_{_W}^2 c_{_W}^2}{ e^2} m_Z^2 \sin^2 2\beta+\bigtriangleup m_h^2,
\label{eq-mh}
\end{eqnarray}
with the main radiative corrections~\cite{ref-mh-rad,ref-mh-rad1}
\begin{eqnarray}
\bigtriangleup m_h^2 = \frac{3m_t^4}{4\pi^2\upsilon^2}\ln \frac{M_S^2}{m_t^2} + \frac{3m_t^4}{4\pi^2\upsilon^2} \frac{X_t^2}{M_S^2} (1-\frac{X_t^2}{12M_S^2}),
\end{eqnarray}
where $\upsilon=174$ GeV, $M_S =\sqrt{m_{{\tilde t}_1}m_{{\tilde t}_2}}$ with $m_{{\tilde t}_{1,2}}$ being the stop masses, $X_t \equiv A_t-\mu\cot\beta$ with $A_t=A_{u_3}$ denoting the trilinear Higgs-stop coupling and $\mu=3\lambda \upsilon_{\nu^c}$ being the Higgsino mass parameter.

Through the analysis of the parameter space in Ref.~\cite{mnSSM1}, we could choose the reasonable values for some parameters as $\kappa=0.4$, $\lambda=0.1$, $A_\lambda=500\;{\rm GeV}$, $\upsilon_{\nu^c}=1\;{\rm TeV}$ and $m_L=m_E={A_{e}}=1\;{\rm TeV}$ for simplicity in the following numerical calculation. Through Eq.~(\ref{tachyon}), we could choose ${A_{\kappa}}=-300\;{\rm GeV}$ to avoid the tachyons.  For the Majorana masses of the gauginos, we will imply the approximate GUT relation $M_1=\frac{\alpha_1^2}{\alpha_2^2}M_2\approx 0.5 M_2$ and $M_3=\frac{\alpha_3^2}{\alpha_2^2}M_2\approx 2.7 M_2$. The gluino mass, $m_{{\tilde g}}\approx M_3$, is larger than about $1.2$ TeV from the ATLAS and CMS experimental data~\cite{ATLAS-sg1,ATLAS-sg2,CMS-sg1,CMS-sg2}. So, we conservatively choose $M_2=1\;{\rm TeV}$. And the first two generations of squarks are strongly constrained by direct searches at the LHC~\cite{ATLAS-sq1,CMS-sq1}. Therefore, we take $m_{{\tilde Q}_{1,2}}=m_{{\tilde u}^c_{1,2}}=m_{{\tilde d}^c_{1,2}}=2\;{\rm TeV}$. The third generation squark masses are not constrained by the LHC as strongly as the first two generations, and affect the SM-like Higgs mass. So, we could adopt $m_{{\tilde Q}_3}=m_{{\tilde u}^c_3}=m_{{\tilde d}^c_3}=1\;{\rm TeV}$. For simplicity, we take $A_{d_{1,2,3}}=A_{u_{1,2}}=1\;{\rm TeV}$. Then, through Eq.~(\ref{eq-mh}), we can choose $\tan \beta=3$ and $A_{u_3}=1.13\;{\rm TeV}$, to keep the SM-like Higgs mass $m_h\simeq 125.7\,{\rm GeV}$.

\begin{figure}
\setlength{\unitlength}{1mm}
\centering
\includegraphics[width=3.1in]{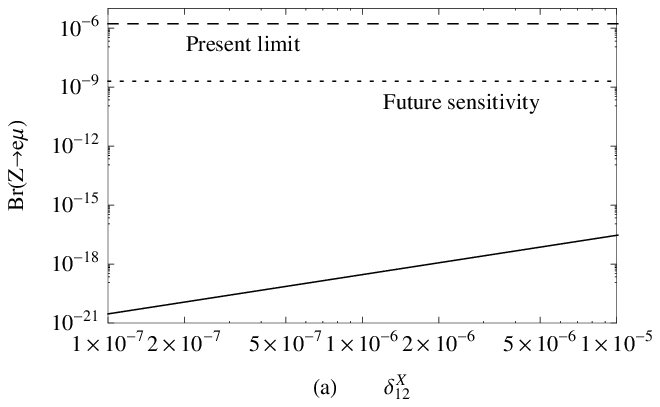}%
\includegraphics[width=3.1in]{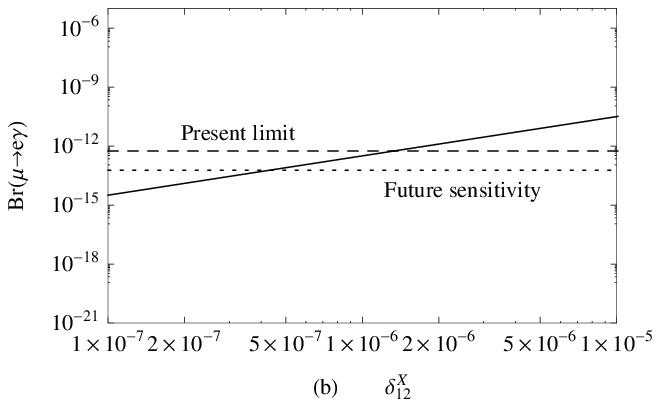}
\includegraphics[width=3.1in]{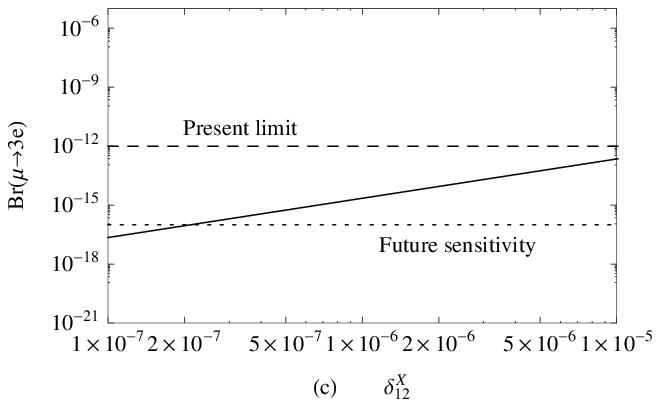}%
\includegraphics[width=3.1in]{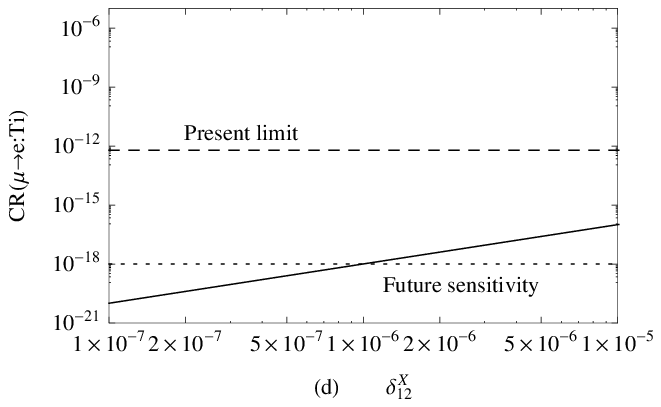}
\caption[]{LFV rates for $\mu-e$ transitions versus slepton mixing parameter $\delta_{12}^X$, where the dashed lines denote the present limits and the dotted lines denote the future sensitivities.}
\label{fig2}
\end{figure}

\begin{figure}
\setlength{\unitlength}{1mm}
\centering
\includegraphics[width=3.1in]{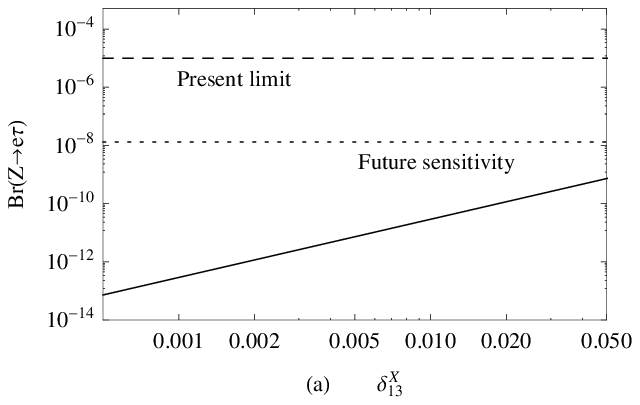}%
\includegraphics[width=3.1in]{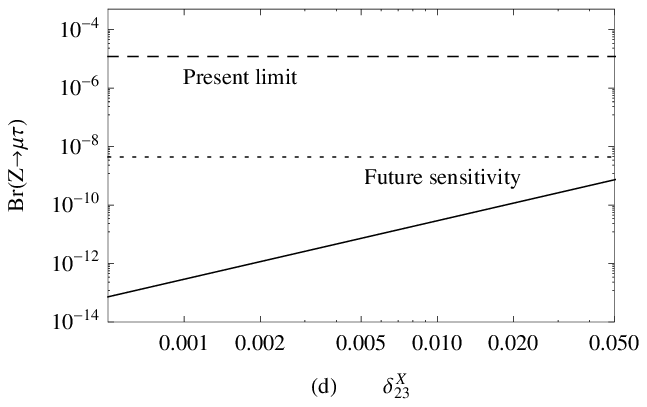}
\includegraphics[width=3.1in]{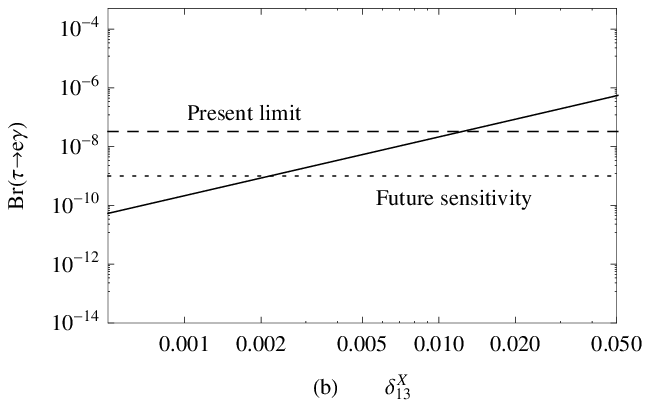}%
\includegraphics[width=3.1in]{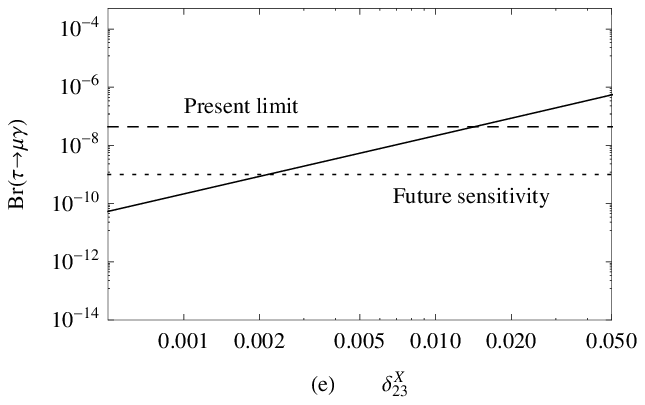}
\includegraphics[width=3.1in]{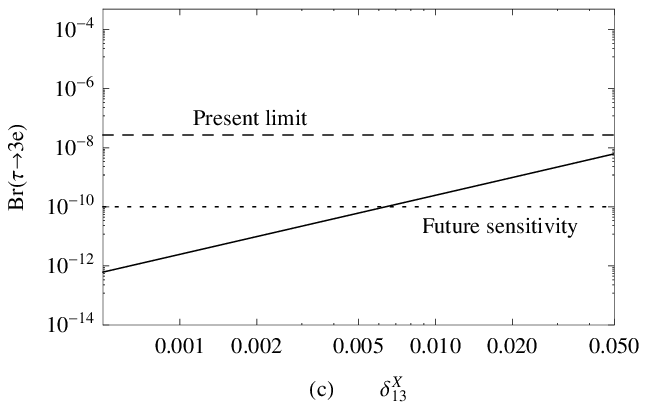}%
\includegraphics[width=3.1in]{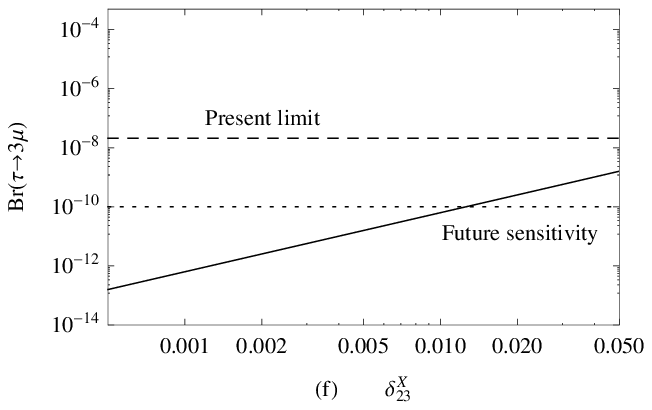}
\caption[]{(a--c) LFV rates for $\tau-e$ transitions versus slepton mixing parameter $\delta_{13}^X$, and (d--f) LFV rates for $\tau-\mu$ transitions versus slepton mixing parameter $\delta_{23}^X$, where the dashed lines denote the present limits and the dotted lines denote the future sensitivities.}
\label{fig3}
\end{figure}

It's well known that the LFV processes are flavor dependent. The LFV rates for $l_j-l_i$  transitions depend respectively on the slepton mixing parameters $\delta_{ij}^X$, which can be numerically confirmed by Fig.~\ref{fig2} and  Fig.~\ref{fig3}. In Fig.~\ref{fig2}, we plot the LFV rates for $\mu-e$ transitions versus slepton mixing parameter $\delta_{12}^X$, where the dashed lines denote the present limits and the dotted lines denote the future sensitivities. Considered that the LFV rates for $\mu-e$ transitions don't depend on $\delta_{13}^X$ and $\delta_{23}^X$, we have chosen $\delta_{13}^X=\delta_{23}^X=0$. The numerical results in Fig.~\ref{fig2} show that the LFV rates for $\mu-e$ transitions are increasing, along with increasing of slepton mixing parameter $\delta_{12}^X$. The branching ratio of $\mu\rightarrow e\gamma$ easily reach the present experimental bound, and constrains $\delta_{12}^X \lesssim \mathcal{O}(10^{-6})$. Under the constraint of the present experimental limit for ${\rm{Br}}(\mu\rightarrow e\gamma)$, the expected rate for $Z\rightarrow e\mu$ still remains orders of magnitude below the future experimental sensitivity, and the expected rates for $\mu\rightarrow 3e$ and $\mu-e$ conversion in nuclei don't reach the present experimental bounds. However, the high future experimental sensitivities still keep a hope to detect $\mu\rightarrow 3e$ and $\mu-e$ conversion in nuclei.
In addition, the dominance of the $\gamma$-mediated channel in the decays $l_j^- \rightarrow l_i^- l_i^- l_i^+$ allows us to derive the simplified relation~\cite{Hisano}
\begin{eqnarray}
\frac{{\rm{Br}}(l_j^- \rightarrow l_i^- l_i^- l_i^+)}{{\rm{Br}}(l_j^- \rightarrow l_i^- \gamma)} \simeq \frac{e^2}{32\pi^2}(\frac{{16}}{3}\ln \frac{{{m_{{l_j}}}}}{{2{m_{{l_i}}}}} - \frac{{14}}{9}),
\end{eqnarray}
which is in agree with the numerical result.

In Fig.~\ref{fig3}(a--c), we picture the LFV rates for $\tau-e$ transitions versus slepton mixing parameter $\delta_{13}^X$, for $\delta_{12}^X=\delta_{23}^X=0$. And we plot the LFV rates for $\tau-\mu$ transitions versus slepton mixing parameter $\delta_{23}^X$ in Fig.~\ref{fig3}(d--f), as $\delta_{12}^X=\delta_{13}^X=0$. The numerical results still show that the LFV rates are increasing, along with increasing of slepton mixing parameters. The present experimental bounds of ${\rm{Br}}(\tau\rightarrow e\gamma)$ and ${\rm{Br}}(\tau\rightarrow \mu\gamma)$ respectively constrain $\delta_{13}^X \lesssim \mathcal{O}(10^{-2})$ and $\delta_{23}^X \lesssim \mathcal{O}(10^{-2})$. Under the constraints of the present experimental limits for ${\rm{Br}}(\tau\rightarrow e\gamma)$ and ${\rm{Br}}(\tau\rightarrow \mu\gamma)$, the expected rates for $Z\rightarrow e\tau$ and $Z\rightarrow \mu\tau$ still remain orders of magnitude below the future experimental sensitivities, and the expected rates for $\tau\rightarrow 3e$ and $\tau\rightarrow 3\mu$ don't reach the present experimental bounds. However, the high future experimental sensitivities still keep a hope to detect $\tau\rightarrow 3e$ and $\tau\rightarrow 3\mu$.

\section{Summary\label{sec5}}

In this paper, we study the LFV processes $Z\rightarrow l_i^\pm l_j^\mp$, $l_j^-\rightarrow l_i^-\gamma$, $l_j^- \rightarrow l_i^- l_i^- l_i^+$ and muon conversion to electron in nuclei with slepton flavor mixing, within framwork of the $\mu\nu$SSM. The numerical results show that the LFV rates for $l_j-l_i$  transitions depend respectively on the slepton mixing parameters $\delta_{ij}^X$, because the LFV processes are flavor dependent. In the $\mu\nu$SSM, the branching ratio of $l_j^-\rightarrow l_i^-\gamma$ can easily reach the present experimental bounds. So, it's a high hope to detect $l_j^-\rightarrow l_i^-\gamma$ in the future. And the present experimental limits of ${\rm{Br}}(l_j^-\rightarrow l_i^-\gamma)$ constrain $\delta_{12}^X \lesssim \mathcal{O}(10^{-6})$, $\delta_{13}^X \lesssim \mathcal{O}(10^{-2})$ and $\delta_{23}^X \lesssim \mathcal{O}(10^{-2})$. Under the constraints of the present experimental limits for ${\rm{Br}}(l_j^-\rightarrow l_i^-\gamma)$, the expected rates for $Z\rightarrow l_i^\pm l_j^\mp$ still remain orders of magnitude below the future experimental sensitivities, and the expected rates for $l_j^- \rightarrow l_i^- l_i^- l_i^+$ and $\mu-e$ conversion in nuclei don't reach the present experimental bounds. However, the high future experimental sensitivities still keep a hope to detect $l_j^- \rightarrow l_i^- l_i^- l_i^+$ and $\mu-e$ conversion in nuclei within the $\mu\nu$SSM.

\begin{acknowledgments}
The work has been supported by the National Natural Science Foundation of China (NNSFC)
with Grant No. 11275036, No. 11047002, the open project of State
Key Laboratory of Mathematics-Mechanization with Grant No. Y3KF311CJ1, the Natural
Science Foundation of Hebei province with Grant No. A2013201277, and Natural Science Fund of Hebei University with Grant No. 2011JQ05, No. 2012-242.
\end{acknowledgments}

\appendix

\section{The couplings\label{app-coupling}}

In this part, we use the indices $i,j=1,\ldots,3$, $\beta,\zeta=1,\ldots,5$, $I=1,\ldots,6$, $\alpha,\rho=1,\ldots,8$ and $\eta=1,\ldots,10$. The couplings of the relative vertices for the LFV processes $Z\rightarrow l_i^\pm l_j^\mp$ in the $\mu\nu$SSM are written by
\begin{eqnarray}
&&\mathcal{L}_{int} =  i C^{Z S_\alpha^- S_\rho^{-\ast}} Z_\mu S_\rho^{-\ast}{\mathord{\buildrel{\lower3pt\hbox{$\scriptscriptstyle\leftrightarrow$}}
\over {\partial^\mu} } } S_\alpha^-  + S_\alpha \bar{\chi}_\zeta (C_L^{{S_\alpha }{\chi _\beta }{{\bar \chi }_\zeta }}{P_L} + C_R^{{S_\alpha }{\chi _\beta }{{\bar \chi }_\zeta }}{P_R}) \chi_\beta   \nonumber\\
&& \qquad\quad + P_\alpha \bar{\chi}_\zeta (C_L^{{P_\alpha }{\chi _\beta }{{\bar \chi }_\zeta }}{P_L}  + C_R^{{P_\alpha }{\chi _\beta }{{\bar \chi }_\zeta }}P_R ) \chi_\beta + S_\alpha^- \bar{\chi}_\beta (C_L^{S_\alpha ^ - \chi _\eta^0 {{\bar \chi }_\beta }}{P_L} + C_R^{S_\alpha ^ - \chi _\eta^0 {{\bar \chi }_\beta }}{P_R} ) \chi_\eta^0  \nonumber\\
&& \qquad\quad + S_\alpha^{-\ast} \bar{\chi}_\eta^0 (C_L^{S_\alpha ^{-\ast} {\chi _\beta }\bar \chi _\eta^0 }{P_L} + C_R^{S_\alpha ^{-\ast} {\chi _\beta }\bar \chi _\eta^0 }{P_R} ) \chi_\beta  + \cdots.
\end{eqnarray}
The coefficients are
\begin{eqnarray}
&&{C^{ZS_\alpha ^ -  S_\rho ^{ -  * }}} = \frac{e}{{2{s_{_W}}{c_{_W}}}}\Big[ ( {1 - 2s_{_W}^2} ){\delta ^{\alpha \rho }} - R{{_{{S^ \pm }}^{(5 + i)\alpha }}^ * }R_{{S^ \pm }}^{(5 + i)\rho } \Big],\\
&&C_L^{{S_\alpha }{\chi _\beta }{{\bar \chi }_\zeta }} = \frac{-e}{{{\sqrt{2}s_{_W}}}}\Big[ R_S^{2\alpha }Z_ - ^{1\beta }Z_ + ^{2\zeta } + R_S^{1\alpha }Z_ - ^{2\beta }Z_ + ^{1\zeta } + R_S^{(2 + i)\alpha }Z_ - ^{(2 + i)\beta }Z_ + ^{1\zeta } \Big]  \nonumber\\
&&\qquad\qquad\quad + \,\frac{1}{\sqrt{2}} {Y_{e_{ij}}}\Big[ R_S^{(2 + i)\alpha }Z_ - ^{1\beta }Z_ + ^{(2 + j)\zeta } - R_S^{1\alpha }Z_ - ^{(2 + i)\beta }Z_ + ^{(2 + j)\zeta }  \Big] \nonumber\\
&&\qquad\qquad\quad - \,\frac{1}{\sqrt{2}}{Y_{\nu_{ij}}}R_S^{(5 + j)\alpha }Z_ - ^{(2 + i)\beta }Z_ + ^{2\zeta } - \frac{1}{\sqrt{2}}{\lambda _i}R_S^{(5 + i)\alpha }Z_ - ^{2\beta }Z_ + ^{2\zeta }  ,\\
&&C_L^{{P_\alpha }{\chi _\beta }{{\bar \chi }_\zeta }} = \frac{{ie}}{{{\sqrt{2}s_{_W}}}}\Big[R_P^{2\alpha }Z_ - ^{1\beta }Z_ + ^{2\zeta } + R_P^{1\alpha }Z_ - ^{2\beta }Z_ + ^{1\zeta } + R_P^{(2 + i)\alpha }Z_ - ^{(2 + i)\beta }Z_ + ^{1\zeta }\Big]  \nonumber\\
&&\qquad\qquad\quad  + \, \frac{i}{\sqrt{2}}{Y_{{e_{ij}}}}\Big[ R_P^{(2 + i)\alpha }Z_ - ^{1\beta }Z_ + ^{(2 + j)\zeta } - R_P^{1\alpha }Z_ - ^{(2 + i)\beta }Z_ + ^{(2 + j)\zeta } \Big] \nonumber\\
&&\qquad\qquad\quad  - \, \frac{i}{\sqrt{2}}{Y_{{\nu _{ij}}}}R_P^{(5 + j)\alpha }Z_ - ^{(2 + i)\beta }Z_ + ^{2\zeta } - \frac{i}{\sqrt{2}}{\lambda _i}R_P^{(5 + i)\alpha }Z_ - ^{2\beta }Z_ + ^{2\zeta } , \\
&&C_L^{S_\alpha^- \chi _\eta^0 {{\bar{\chi}}_\beta }} =   \frac{-e}{{\sqrt{2} {s_W}{c_W}}}R{_{{S^\pm }}^{2\alpha \ast } }Z_+^{2\beta} \Big[ {{c_W}Z_n^{2\eta } + {s_W}Z_n^{1\eta }} \Big]  - \frac{e}{{{s_W}}}R{_{{S^ \pm }}^{2\alpha\ast } }Z_ + ^{1\beta }Z_n^{4\eta }  \nonumber\\
&&\qquad\qquad\quad - \frac{{\sqrt{2} e}}{{{s_W}}}R{_{{S^\pm }}^{(5 + i)\alpha\ast } }Z_ + ^{(2 + i)\beta }Z_n^{1\eta } + {Y_{\nu_{ij}}}R_{{S^ \pm }}^{(2 + i)\alpha }Z_ + ^{2\beta }Z_n^{(4 + j)\eta } \nonumber\\
&&\qquad\qquad\quad  + \, {Y_{e_{ij}}}Z_ + ^{(2 + j)\beta } \Big[ R_{{S^ \pm }}^{1\alpha }Z_n^{(7 + i)\eta } - R_{{S^ \pm }}^{(2 + i)\alpha }Z_n^{3\eta } \Big] - {\lambda _i}R_{{S^ \pm }}^{1\alpha }Z_ + ^{2\beta }Z_n^{(4 + i)\eta },\\
&&C_L^{S_\alpha ^{-\ast} {\chi _\beta }\bar \chi _\eta^0 } =   \frac{e}{{\sqrt 2 {s_W}{c_W}}}\Big[ R{{_{{S^ \pm }}^{1\alpha\ast }} }Z_ - ^{2\beta } + R{{_{{S^ \pm }}^{(2 + i)\alpha }}^ * }Z_ - ^{(2 + i)\beta }\Big]\Big[ {c_W}Z_n^{2\eta } + {s_W}Z_n^{1\eta }\Big] \nonumber\\
&&\qquad\qquad\quad\;\; - \frac{e}{{{s_W}}}Z_ - ^{1\beta }\Big[ R{{_{{S^ \pm }}^{1\alpha\ast }} }Z_n^{3\eta } + R{{_{{S^ \pm }}^{(2 + i)\alpha\ast }} }Z_n^{(7 + i)\eta }\Big] + {Y_{\nu_{ij}}}R_{{S^ \pm }}^{2\alpha }Z_ - ^{(2 + i)\beta }Z_n^{(4 + j)\eta }\nonumber\\
&&\qquad\qquad\quad\;\; +\: {Y_{{e_{ij}}}}R_{{S^ \pm }}^{(5 + j)\alpha }\Big[ Z_ - ^{2\beta }Z_n^{(7 + i)\eta } - Z_ - ^{(2 + i)\beta }Z_n^{3\eta } \Big] - {\lambda _i}R_{{S^ \pm }}^{2\alpha }Z_ - ^{2\beta }Z_n^{(4 + i)\eta },\\
&&C_R^{{S_\alpha }{\chi _\beta }{{\bar \chi }_\zeta }} = \Big[ {C_L^{{S_\alpha }{\chi _\zeta }{{\bar \chi }_\beta }}} \Big]^ * ,\qquad\quad    C_R^{{P_\alpha }{\chi _\beta }{{\bar \chi }_\zeta }} = \Big[ {C_L^{{P_\alpha }{\chi _\zeta }{{\bar \chi }_\beta }}} \Big]^ * ,\\
&&C_R^{S_\alpha ^ - \chi _\eta^0 {{\bar \chi }_\beta }} = \Big[ {C_L^{S_\alpha ^{-\ast} {\chi _\beta }\bar \chi _\eta^0 }} \Big]^ * , \qquad  C_R^{S_\alpha ^{-\ast} {\chi _\beta }\bar \chi _\eta^0 } =\Big[ {C_L^{S_\alpha ^ - \chi _\eta^0 {{\bar \chi }_\beta }}} \Big]^ * ,
\end{eqnarray}
where $R_{S}$, $R_{P}$, $R_{S^\pm}$, $Z_{\mp}$ and $Z_n$ can be found in Ref.~\cite{ref-zhang}.

\end{document}